\documentclass[aps,amsmath,amssymb,preprintnumbers,groupedaddress,twocolumn,floatfix,nofootinbib]{revtex4-1}            
\usepackage{amsmath}
\usepackage{amsfonts}
\usepackage{amssymb}
\usepackage{stmaryrd}
\usepackage{empheq}
\usepackage[arrow,matrix,curve]{xy}

\usepackage{graphicx} 
\usepackage{pstricks}
\usepackage{hyperref}

\begin{document}


\preprint{MPP-2023-193}

\title{The Emergent M-theory Limit}

\author{Ralph Blumenhagen$^{1,2}$} 
\author{Niccol\`o Cribiori$^{1}$}
\author{Aleksandar Gligovic$^{1,2}$}
\author{Antonia Paraskevopoulou$^{1,3}$}

\affiliation{
$^{1}$ Max-Planck-Institut f\"ur Physik, F\"ohringer Ring 6, 80805 M\"unchen, Germany\\
$^{2}$ Exzellenzcluster ORIGINS, Boltzmannstr. 2, D-85748 Garching, Germany\\
$^{3}$ Ludwig-Maximilians-Universit{\"a}t M\"unchen, Fakult{\"a}t f{\"u}r Physik, Theresienstr.~37, 80333 M\"unchen, Germany
}


\begin{abstract}
It has been recently proposed that at each infinite distance limit in the moduli space of quantum gravity a perturbative description emerges with fundamental degrees of freedom given by those infinite towers of states whose typical mass scale is parametrically not larger than the ultraviolet cutoff, identified with the species scale. 
This proposal is applied to the familiar ten-dimensional type IIA and IIB superstring theories, when considering the limit of infinite string coupling.
For type IIB, the light towers of states are given by excitations of  the $D1$-brane, as expected from self-duality. 
Instead, for type IIA at strong coupling, which is dual to M-theory on $S^1$, we make the observation that the emergent degrees of freedom are bound states of transversal $M2$- and $M5$-branes with Kaluza-Klein momentum along the circle. We speculate on the interpretation of the necessity of including all these states for a putative quantum formulation of M-theory.
\end{abstract}



\maketitle




\section{Introduction}
\label{sec:intro}

The swampland program \cite{Vafa:2005ui}, see e.g.~\cite{Palti:2019pca, vanBeest:2021lhn,Agmon:2022thq}
for reviews, aims at providing a broader perspective on the rules governing quantum gravity, transcending previous heroic efforts in concrete model building under the string lamppost. 
In this respect, there appear to exist tight constraints on which low energy effective action can be completed into a full theory of quantum gravity in the ultraviolet. In turn, this gives rise to new naturalness principles with potentially deep implications.
These ideas are powerful enough to address some of the most important questions in contemporary Theoretical Physics, such as the cosmological constant problem, dark matter or the mechanism for inflation.

In this letter, we apply certain swampland criteria to investigate the notoriously mysterious quantum nature of M-theory, which has been proposed to be the missing corner relating all superstrings in ten dimensions via dualities \cite{Witten:1995ex}. 
Some aspects of the theory are known, as the fact that its low-energy action is eleven-dimensional supergravity, and that it contains $M2$-branes and $M5$-branes, which are respectively electrically and magnetically charged under a three-form gauge field. 
There have been approaches to define a quantum version in terms of a matrix theory or, similarly to string theory, by a first quantization of the membrane action \cite{deWit:1988wri}, see \cite{Taylor:2001vb} for a review. 
However, as of today it is fair to say that the complete picture has not been understood. 
In the present work, we shed new light on this open problem.

Our approach is essentially based on the Swampland Distance Conjecture \cite{Ooguri:2006in} and on one of its latest developments, namely the Emergent String Conjecture \cite{Lee:2019wij}.
In \cite{BCGP} we pushed these ideas one step further and proposed that at each infinite distance limit in moduli space with the vacuum expectation value of a modulus  $t_0=\langle t\rangle$ taken to infinity, a perturbative quantum gravity theory emerges with a characteristic set of fundamental degrees of freedom.
They consist of the full infinite towers of states whose typical mass scale is parametrically not larger than the species scale, taken to be the
ultraviolet cutoff \cite{Dvali:2007hz}. 
Note that this recipe to probe quantum gravity is conceptually different from the usual definition of effective field theories having dynamical degrees of freedom strictly below an ultraviolet cutoff.
In addition, the novel theory probed in this way was claimed to admit a perturbative expansion in the inverse of the species number, $g_E\simeq t_0^{-1}\simeq N_{\rm sp}^{-1}$, which naturally provides a small parameter.

When testing this proposal within the vector multiplet moduli space of
compactifications of type IIA string theory on a Calabi-Yau threefold
$X$, in \cite{BCGP} we related the decompactification limit,
engineered in such a way to keep the four-dimensional Planck scale
fixed, to the computation of Gopakumar--Vafa invariants
\cite{Gopakumar:1998ii,Gopakumar:1998jq}.
Indeed, in this limit the new perturbative states below the species scale turned out to be $D0$/$D2$/$NS5$-brane bound states. 
Surprisingly, they are not just particle-like, but contain also excitations of $h_{11}(X)$ emergent strings from the $NS5$-branes.
From the dual point of view, this corresponds to M-theory on $X\times S^1$ in the limit of large circle, leading to decompactification to five-dimensions.
In the same setup, an emergent string limit \cite{Lee:2019wij} was instead better understood via the well-known heterotic/IIA duality and hence described in terms of the dual weakly coupled string becoming asymptotically tensionless.

It is a natural question how to apply the same approach directly to higher dimensional theories, such as the type II superstrings in ten dimensions.
The relevant modulus is the dilaton, $g_s=\exp(\phi )$,  and it admits two infinite distance limits, namely small and large string coupling. 
In this context, the quantum theory approached at the limit is so far better understood when the new fundamental degrees of freedom are excitations of a single emergent string \cite{Lee:2019wij}. 
This will be the case for the strong coupling limit of the type IIB superstring, thus confirming that the proposal of \cite{BCGP} is in accordance with well-known facts about dualities. 
The strong coupling limit of type IIA, which is the main focus of the present letter, will instead be richer and will provide new information on a putative quantum formulation of M-theory on a circle.

\section{Emerging Theories}

We focus on the quantum gravity theories emerging in the strong coupling limit, $g_s\to\infty$, of the ten-dimensional type IIA and IIB superstrings.
According to the recent proposal of \cite{BCGP}, in such an infinite distance limit in moduli space we expect to get a new perturbative description of the theory, for which the fundamental degrees of freedom are given by those infinite towers of states whose typical mass scale is parametrically not larger than the corresponding species scale, assumed to be the ultraviolet cutoff.

Besides the tension of the various BPS branes in type IIA and IIB, there are only a few more relations we need for our purposes. 
First, we have to clarify what we mean by typical mass scale of a tower, denoted with $\frak m$ in what follows. As explained in more detail in \cite{BCGP}, we associate to a BPS $p$-brane of tension $T_p$ the natural mass scale
\begin{equation}
\label{Deltamp}
\frak m_p\sim T_p^{\frac{1}{p+1}}\,.
\end{equation}
We understand that ${\frak m}_p$ reflects the mass levels in the tower of particle-like excitations of the $p$-brane.
This is certainly true for BPS $D0$-branes and for the excitations
of the perturbative fundamental string. In other cases, since we do not know how to quantize the $p$-brane from first principles, we consider it
as our main working assumption.
We remark that if a proper quantization of the $p$-brane turns out to lead to a large number of excitations at a scale parametrically smaller than \eqref{Deltamp}, our estimate of the species scale needs to be revisited.

We will also need the relation between the Planck scale and the fundamental string scale, which in ten dimensions reads
\begin{equation}
  M_{\rm pl}={M_s\over g_s^{1/4}}\,,
\end{equation}
while that between the species scale $ \Lambda$ and the number of light species $N_{\rm sp}$ is \cite{Dvali:2007hz}
\begin{equation}
\label{speciescale}
\Lambda={M_{\rm pl}\over N_{\rm sp}^{1/8}}\,.
\end{equation}    

In the following, we look at the strong coupling limit of type II
superstrings and investigate which new light degrees of freedom
arise.

\subsection{Strong coupling limit of type IIB}

Let us start with the type IIB superstring and take $g_{s} \to \infty$. 
The lightest tower of states in this limit are excitations of a $D1$-string with tension $T_{D1}=M_s^2/g_s=M_{\rm pl}^2/g_s^{1/2}$. 
The species scale for a ($D1$-)string is given by the ($D1$-)tension, i.e. $ \Lambda^2_{\rm IIB} = T_{D1}$, hence we have
\begin{equation}
\label{IIBspeciesscale}
\Lambda_{\rm IIB}={M_{\rm pl}\over g_s^{1/4}}\,.
\end{equation}

As proposed in \cite{BCGP}, in the strong coupling regime, $g_s\gg 1$, we introduce a new small parameter $g_E=1/N_{sp}=1/g^2_s$, which can be used to organize a perturbative expansion of the emergent theory. 
Then, we go through the list of all other type IIB branes and express the corresponding towers of excitations in terms of the species scale and the coupling $g_E$. 
The goal is to identify all those towers whose typical mass scale  is parametrically not larger than the species scale. 
These are the new perturbative degrees of freedom of the emergent theory. 

For the fundamental string, we find
\begin{equation}
\frak m_{F1}= M_{s}  = \frac{ \Lambda_{\rm IIB}}{g_E^{1/4}}\,,
\end{equation}
For the $NS5$-brane, we get
\begin{equation}
\frak m_{NS5}= {M_s\over g_s^{1/3}} = \frac{ \Lambda_{\rm IIB}}{g_E^{1/12}}\,, 
\end{equation} 
while for the type IIB $Dp$-branes ($p\geq 1$), we can write collectively
\begin{equation}
\frak m_{Dp} = \frac{ \Lambda_{ \rm IIB}}{(g_E)^{\beta}}\,, 
\end{equation}
with $\beta = \frac{1}{4} \frac{p-1}{p+1}$. From these expressions,
one can see that all excitations of the fundamental string, of the
$NS5$-brane and of all $Dp$-branes with $p>1$ are parametrically
heavier than the  species scale \eqref{IIBspeciesscale} in the regime $g_E \ll 1$. 

Hence, we conclude that the theory emerging in the strong coupling limit of type IIB is again a string theory. 
The new perturbative states are given by the excitations of the
$D1$-brane, with ${\frak m}_{D1} =  \Lambda_{\rm IIB}$. 
All other objects have excitations of typical mass scale parametrically larger than the species scale and are thus considered non-perturbative.
These findings are of course consistent with the well-established self-duality of the type IIB superstring, which maps the strongly coupled $D1$-brane to the weakly coupled fundamental string. 
In fact, the perturbative expansion in $\sqrt{g_E}$ arises as a sum over all topological configurations of a two-dimensional string, which is the familiar genus expansion of string theory.
One can view this as a non-trivial consistency check for the proposal of \cite{BCGP}.

\subsection{Strong coupling limit of type IIA}

Let us now perform the same analysis for the type IIA superstring.
It has been proposed that, for $g_s\to \infty $, the theory is dual to M-theory on a circle of radius \cite{Witten:1995ex}
\begin{equation}
r_{11} = R_{11} M_*\,,\qquad g_s=(r_{11})^{3/2}\,.
\end{equation}
Here, $M_*$ is the eleven-dimensional Planck scale and $R_{11}$ the size of the eleventh direction. The ten-dimensional Planck scale is given as
\begin{equation}
M_{\rm pl}^8 = (M_*)^9 R_{11}\,.
\end{equation}

The lightest states in the limit are  BPS bound states of
$D0$-branes. Their mass scale  is
\begin{equation}
\frak m_{D0} = \frac{M_s}{g_s} = \frac{M_{\rm pl}}{g_s^{3/4}}= \frac{M_*}{r_{11}}\,.
\end{equation}  
These states are particle-like and contribute a number of light species $N_{\rm sp}=\Lambda_{\rm IIA}/\frak m_{D0}$.
Using \eqref{speciescale}, one can then determine the type IIA species scale as
\begin{equation}
\Lambda_{\rm IIA}= {M_{\rm pl}\over g_s^{1/12}}=M_*\,.
\end{equation}                
As expected for decompactification limits, the species scale is given by the higher dimensional Planck scale. Notice that $ \Lambda_{\rm IIB} \neq  \Lambda_{\rm IIA}$, which explains the different behavior of the two theories in the strong coupling limit.

Contrary to the type IIB case, the mass spacing of the lightest tower of states is now parametrically smaller than the species scale. Therefore, there might exist other towers with mass spacing between $\frak m_{D0}$ and $ \Lambda_{\rm IIA}$. According to \cite{BCGP}, they must be included as degrees of freedom of the emergent theory as well.

To test this possibility, we introduce the coupling constant $g_E =1/N_{sp}=1/g_s^{2/3}= 1/r_{11}$ and we go through the list of type IIA objects.
For the fundamental string, we find
\begin{equation}
\frak m_{F1} = \frac{ \Lambda_{\rm IIA}}{g_E^{1/2}} .
\end{equation}
For the $NS5$-brane, we get
\begin{equation}
\frak m_{NS5} =  \Lambda_{\rm IIA}.
\end{equation} 
For the type IIA $Dp$-branes, we can write collectively
\begin{equation}
  \frak m_{Dp} = \frac{ \Lambda_{\rm IIA}}{(g_E)^{\alpha}}, 
\end{equation}
with $\alpha = \frac{1}{2} \frac{p-2}{p+1}$.
From these expressions, one can see that bound states of
$D0/D2/NS5$-branes, together with their excitations, have a typical
mass scale  not
larger than the species scale. Hence, they are emergent degrees of freedom.
 Instead, the other objects give rise to heavier towers, non-perturbative in $g_E$ and with a mass spacing parametrically larger than the species scale. 
From the dual M-theory point of view, $D2$-branes and $NS5$-branes are dual to transversal (to $S^1$) $M2$-branes and $M5$-branes respectively. The other type IIA branes are all wrapping the large $S^1$, explaining why they are heavier in the limit. Notice also that this picture is different from the usual experience with string theories, since here electric $M2$-branes and magnetic dual $M5$-branes give rise to excitations with the same mass.

\subsection{Speculations on M-theory}

To summarize, our findings indicate that three objects have to be included in the perturbative quantum gravity theory emerging in the strong coupling limit of the type IIA superstring. 
These are $D0$-, $D2$- and $NS5$-branes or, from the dual M-theory perspective, transversal $M2$- and $M5$-branes carrying discrete Kaluza-Klein momentum along the $S^1$ with size $R_{11}$. 
This suggests that any attempt to include only parts of these states, as for example only $D0$-branes or only the excitations of the membrane, is doomed to be incomplete, unless the missing degrees of freedom somehow emerge, as it was e.g.~proposed in the BFSS matrix-model \cite{Banks:1996vh}.

\begin{figure}[ht]
\centering
\includegraphics[scale=0.28]{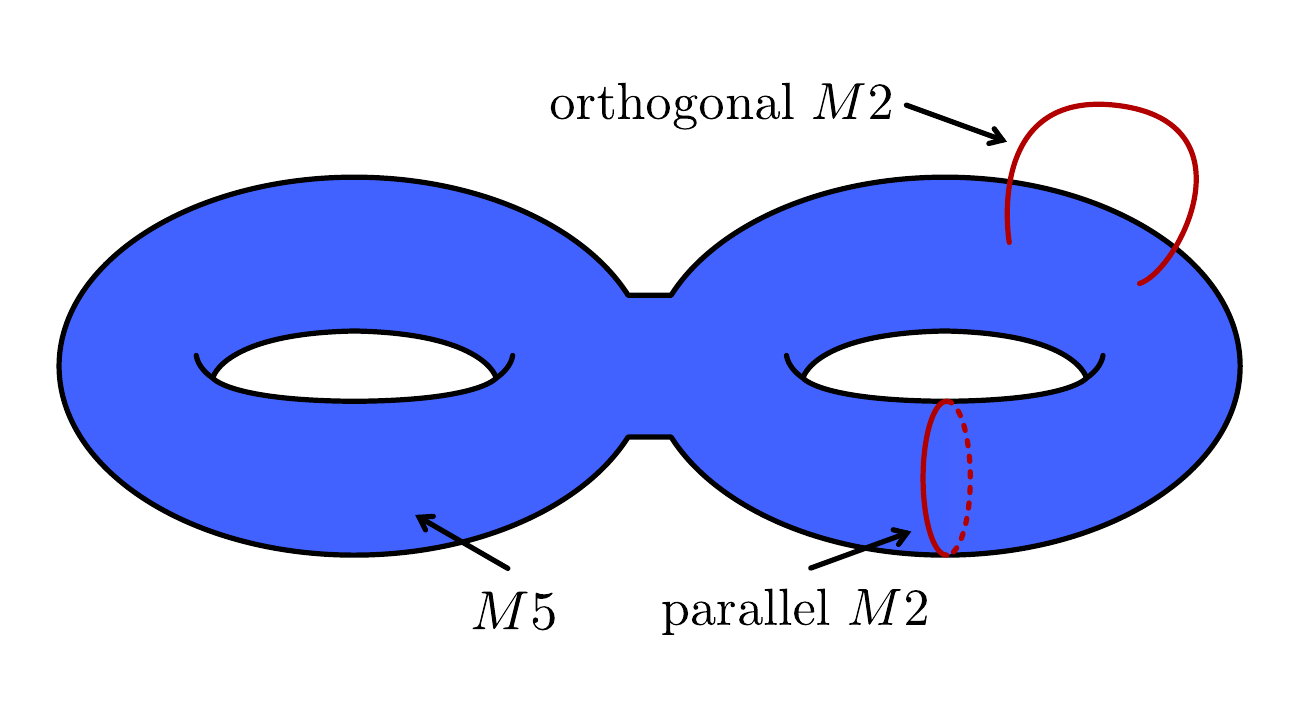}
\caption{$M5$-brane with $M2$-branes ending on it. The $(2=6/3)$-dimensional analogue of the actual situation is shown.}
\label{fig:M2M5}
\end{figure}

For instance, it has always been puzzling how a theory of only membranes could give rise to something like a genus expansion (as for strings), for the Euler characteristic of a 3-manifold always vanishes. 
By including the $M5$-branes from the start, which have an even-dimensional world-volume, one is led to consider bound states of $M5$/$M2$-branes where such a genus expansion might exist, as figure \ref{fig:M2M5} suggests.

From the type IIA perspective, it is known that there are bound states of parallel $D0/D2/NS5$-branes \cite{Mitra:2000wr}. 
However a $D2$-brane can also end on an $NS5$-brane, being charged under the self-dual 2-form living on the latter and breaking $1/2$ of the supercharges.
In analogy to the type IIB case, we can speculate that when dealing with M-theory in the perturbative limit of a large eleventh direction ($S^1$), one needs to sum over all topological configurations of $M2/M5$-branes bound states. 
Since the eleventh direction is only felt by the Kaluza-Klein modes, the perturbative expansion in $g_E=1/r_{11}$  does not seem to be related to this sum over $M2/M5$ bound states, so that the latter might prevail also when the circle decompactifies. 
Then, all former non-perturbative states in $g_E$, i.e.~the longitudinal branes, become infinitely massive and one is left with only the perturbative $M2/M5$-brane states.

Upon compactification of M-theory on a Calabi-Yau threefold times a circle, an $M5$-brane can wrap a divisor $D$ giving rise to an asymptotically tensionless string in the non-compact four dimensions.
These are precisely the type IIA NS5-branes discussed in \cite{BCGP}.
Then, the genus expansion of this emerging string lifts to a family of $M5$-brane geometries of the type $D\times C_g$, where $C_g$ is a Riemann surface of genus $g$. This is reminiscent of the set-up of the AGT correspondence \cite{Alday:2009aq}.

For classical on-shell configurations, the world-volume theory of such strings is governed by a supersymmetric CFT in two-dimensions. 
Hence, it is tempting to speculate that it uplifts to a maximally supersymmetric $(0,2)$ SCFT on the six-dimensional word-volume of the $M5$-brane. 
Allowing a theory with $M2$-branes ending on the $M5$-branes would break this to a $(0,1)$ SCFT.

We would like to conclude by stressing that to confirm our proposal further evidence is needed, in particular with respect to our central assumption, namely that a proper quantization of the world-volume theories on the $M2$ and the $M5$ does not lead to a large number of states below their typical mass scales, spoiling the prediction of the species scale based on $D0$-branes.

\section{Summary}

In this letter, we formulated a proposal for a putative perturbative description of M-theory (on a circle). Concretely, we suggested that the light degrees of freedom are given by excitations of transversal (to the circle) $M2/M5$-branes carrying Kaluza-Klein momentum along the compactified eleventh direction.

Our proposal is based on recent developments in the swampland program. 
In particular, it follows from applying the general strategy of \cite{BCGP} to the moduli space of the critical type II superstrings in the strong coupling limit. 
While for the type IIB case we recover the result known from self-duality, namely that at strong coupling a perturbative string theory emerges with light degrees of freedom given by excitation of the $D1$-brane, in type IIA the situation is richer. 
Indeed, at strong coupling we find evidence for the existence of a perturbative theory of quantum gravity with degrees of freedom given by bound states of $D0/D2/NS5$-branes, together with their excitations. 
By applying type IIA/M-theory duality, we arrive at the main result of
our letter, namely that a perturbative description of M-theory in the
coupling $1/r_{11}$ should be given by transversal $M2/M5$-branes
bound states carrying Kaluza-Klein momentum along the eleventh
direction.

\begin{figure}[h]
\centering
\includegraphics[scale=0.25]{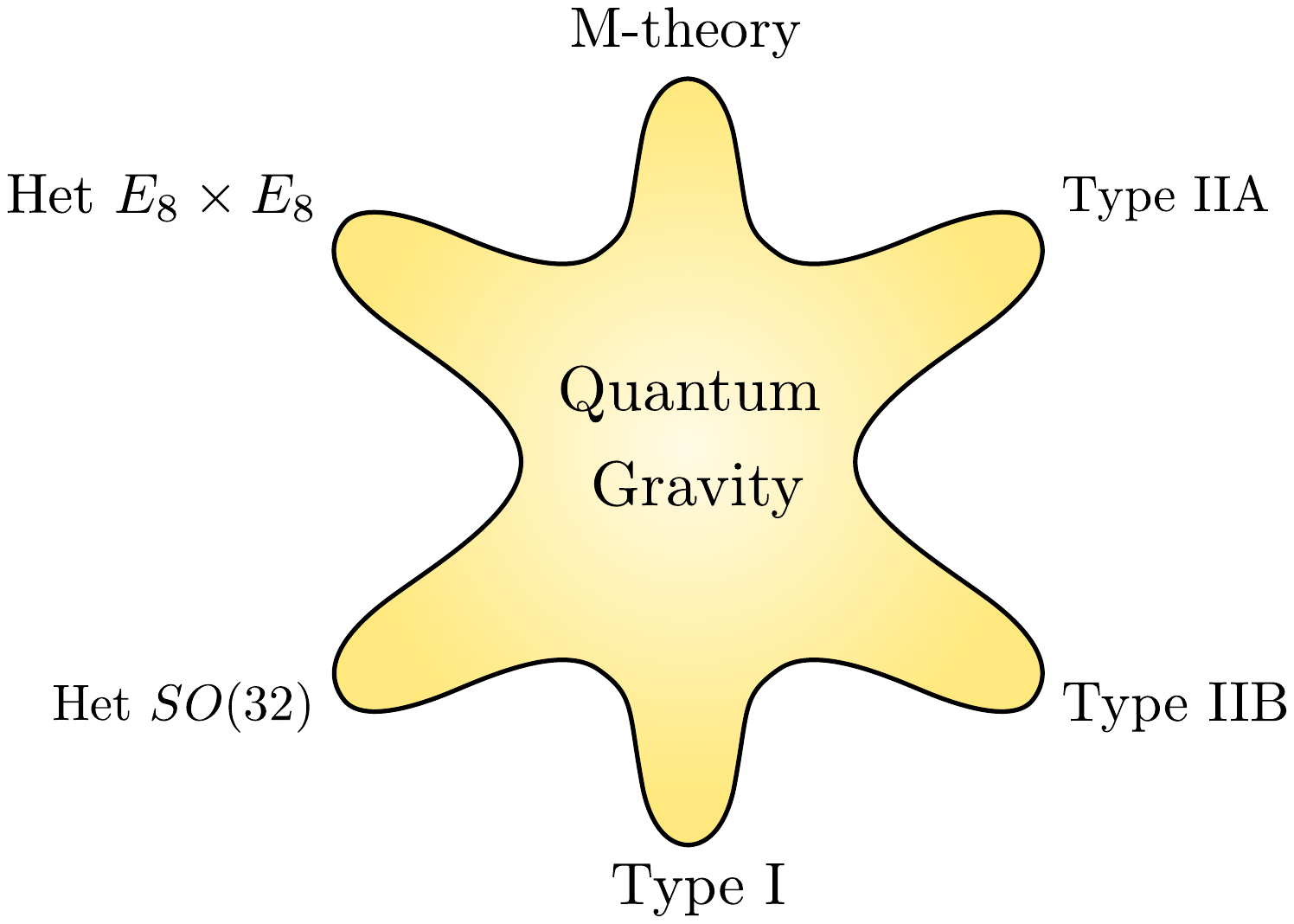}
\caption{The duality star: each corner corresponds to a perturbative theory of quantum gravity.}
\label{fig:dualstar}
\end{figure}

As shown in figure \ref{fig:dualstar}, this theory occupies the final perturbative spike of the duality star, with the other ones given by the familiar (perturbative) critical superstring theories in ten dimensions. 
For the even more elusive and genuinely non-perturbative quantum gravity
theory  in the middle,  the  mass gap will be of order the
ten-dimensional Planck scale and all couplings of order one.


\begin{acknowledgments}
We thank Timo Weigand for helpful discussions. The work of R.B. and A.G. is funded by the Deutsche Forschungsgemeinschaft (DFG, German Research Foundation) under Germany’s Excellence Strategy – EXC-2094 – 390783311. 
The work of N.C.~is supported by the Alexander-von-Humboldt foundation. 
\end{acknowledgments}




\begin{thebibliography}{0}%
\makeatletter
\providecommand \@ifxundefined [1]{%
 \@ifx{#1\undefined}
}%
\providecommand \@ifnum [1]{%
 \ifnum #1\expandafter \@firstoftwo
 \else \expandafter \@secondoftwo
 \fi
}%
\providecommand \@ifx [1]{%
 \ifx #1\expandafter \@firstoftwo
 \else \expandafter \@secondoftwo
 \fi
}%
\providecommand \natexlab [1]{#1}%
\providecommand \enquote  [1]{``#1''}%
\providecommand \bibnamefont  [1]{#1}%
\providecommand \bibfnamefont [1]{#1}%
\providecommand \citenamefont [1]{#1}%
\providecommand \href@noop [0]{\@secondoftwo}%
\providecommand \href [0]{\begingroup \@sanitize@url \@href}%
\providecommand \@href[1]{\@@startlink{#1}\@@href}%
\providecommand \@@href[1]{\endgroup#1\@@endlink}%
\providecommand \@sanitize@url [0]{\catcode `\\12\catcode `\$12\catcode
  `\&12\catcode `\#12\catcode `\^12\catcode `\_12\catcode `\%12\relax}%
\providecommand \@@startlink[1]{}%
\providecommand \@@endlink[0]{}%
\providecommand \url  [0]{\begingroup\@sanitize@url \@url }%
\providecommand \@url [1]{\endgroup\@href {#1}{\urlprefix }}%
\providecommand \urlprefix  [0]{URL }%
\providecommand \Eprint [0]{\href }%
\providecommand \doibase [0]{http://dx.doi.org/}%
\providecommand \selectlanguage [0]{\@gobble}%
\providecommand \bibinfo  [0]{\@secondoftwo}%
\providecommand \bibfield  [0]{\@secondoftwo}%
\providecommand \translation [1]{[#1]}%
\providecommand \BibitemOpen [0]{}%
\providecommand \bibitemStop [0]{}%
\providecommand \bibitemNoStop [0]{.\EOS\space}%
\providecommand \EOS [0]{\spacefactor3000\relax}%
\providecommand \BibitemShut  [1]{\csname bibitem#1\endcsname}%
\let\auto@bib@innerbib\@empty
\end{thebibliography}%


\begin{thebibliography}{9}

\bibitem{Vafa:2005ui}
C.~Vafa,
[arXiv:hep-th/0509212 [hep-th]].

\bibitem{Palti:2019pca}
E.~Palti,
Fortsch. Phys. \textbf{67}, no.6, 1900037 (2019)


\bibitem{vanBeest:2021lhn}
M.~van Beest, J.~Calder\'on-Infante, D.~Mirfendereski and I.~Valenzuela,
Phys. Rept. \textbf{989} (2022), 1-50


\bibitem{Agmon:2022thq}
N.~B.~Agmon, A.~Bedroya, M.~J.~Kang and C.~Vafa,
[arXiv:2212.06187 [hep-th]].

\bibitem{Witten:1995ex}
E.~Witten,
Nucl. Phys. B \textbf{443} (1995), 85-126


\bibitem{deWit:1988wri}
B.~de Wit, J.~Hoppe and H.~Nicolai,
Nucl. Phys. B \textbf{305} (1988), 545


\bibitem{Taylor:2001vb}
W.~Taylor,
Rev. Mod. Phys. \textbf{73}, 419-462 (2001)


\bibitem{Ooguri:2006in}
H.~Ooguri and C.~Vafa,
Nucl. Phys. B \textbf{766}, 21-33 (2007)

\bibitem{Lee:2019wij}
S.~J.~Lee, W.~Lerche and T.~Weigand,
JHEP \textbf{02}, 190 (2022)


\bibitem{BCGP}
R.~Blumenhagen, N.~Cribiori, A.~Gligovic and A.~Paraskevopoulou,
[arXiv:2309.xxxxx [hep-th]].




\bibitem{Dvali:2007hz}
G.~Dvali,
Fortsch. Phys. \textbf{58} (2010), 528-536


\bibitem{Gopakumar:1998ii}
R.~Gopakumar and C.~Vafa,
[arXiv:hep-th/9809187 [hep-th]].

\bibitem{Gopakumar:1998jq}
R.~Gopakumar and C.~Vafa,
[arXiv:hep-th/9812127 [hep-th]].



\bibitem{Banks:1996vh}
T.~Banks, W.~Fischler, S.~H.~Shenker and L.~Susskind,
Phys. Rev. D \textbf{55} (1997), 5112-5128


\bibitem{Mitra:2000wr}
I.~Mitra and S.~Roy,
JHEP \textbf{02}, 026 (2001)


\bibitem{Alday:2009aq}
L.~F.~Alday, D.~Gaiotto and Y.~Tachikawa,
Lett. Math. Phys. \textbf{91}, 167-197 (2010)



\end{thebibliography}
\end{document}